\documentclass[pdflatex,sn-mathphys-num]{sn-jnl}% Math and Physical Sciences Numbered Reference Style
\usepackage{graphicx} % 
\usepackage{subcaption}
\usepackage{amsmath} 
\usepackage{braket}
\usepackage{multirow}%
\usepackage{amsmath,amssymb,amsfonts}%
\usepackage{amsthm}%
\usepackage{mathrsfs}%
\usepackage[title]{appendix}%
\usepackage{xcolor}%
\usepackage{textcomp}%
\usepackage{manyfoot}%
\usepackage{booktabs}%
\usepackage{algorithm}%
\usepackage{algorithmicx}%
\usepackage{algpseudocode}%
\usepackage{listings}%
\usepackage[normalem]{ulem}
\usepackage{eso-pic} %for the preprint number
\newcommand{\cO}{ {\cal O} }

\theoremstyle{thmstyleone}%
\theoremstyle{thmstyletwo}%

\theoremstyle{thmstylethree}%
\begin{document}
%%%%%%%%%%%%%%%%%%%%%%%%%%%%%%%%%%%%%%%%%%%%%%%%
\AddToShipoutPictureFG*{%
  \AtPageUpperLeft{%
    \put(\LenToUnit{\dimexpr\paperwidth-2cm},\LenToUnit{\dimexpr-1.5cm}){%
      \makebox(0,0)[r]{\small\texttt{Preprint  IPARCOS-UCM-26-059}}%
    }%
  }%
}

\title{Efficient Iterative Symmetrization
algorithm for quantum computing and hadron applications 
}


\author[1]{\fnm{Nahia J.} \sur{Dios-Bilbao}}

\author[1]{\fnm{Adrián} \sur{Castaño García}}

\author*[1]{\fnm{J. J.} \sur{G\'alvez-Viruet}}\email{juagalve@ucm.es}

\author[1]{\fnm{Felipe J.} \sur{Llanes-Estrada}}

\author[1]{\fnm{Mario} \sur{Logrosán-Álvarez}}

\affil*[1]{\orgdiv{F\'{i}sica Te\'orica \& IPARCOS}, \orgname{Universidad Complutense de Madrid}, \orgaddress{\street{Fac. de CC. Físicas, Plaza de las Ciencias 1}, \city{Madrid}, \postcode{28040},  \country{Spain}}}

\author[2]{\fnm{Nicolás M.} \sur{Arenaza}}

\affil[2]{\orgdiv{Instituto de F\'{i}sica Te\'orica }, \orgname{C/ Nicolás Cabrera 13-15, Universidad Autónoma de Madrid}, %\orgaddress{\street{Plaza de las Ciencias 1}, 
\city{Madrid}, \
\postcode{28049},  \country{Spain}}

%\affil[2]{\orgdiv{Department}, \orgname{Organization}, \orgaddress{\street{Street}, \city{City}, \postcode{10587}, \state{State}, \country{Country}}}

%%%%%%%%%%%%%%%%%%%%%%%%%%%%%%%%%%%%%%%%%%%%%%%
\abstract{Efficient algorithms for the (anti)symmetrization of quantum registers are necessary for upcoming few- and many-body physics applications, from atomic to particle-physics scales. We propose a recursive algorithm  (anti)symmetrizing the last particle added to memory, each time one is created or destroyed, respect to the rest of already (anti)symmetrized ones.
Given a generalized Dicke state (GDS) of $(j-1)$ quantum registers, each assigned to a particle, the algorithm produces another GDS of $j$ registers (of $n_{qb}$ qubits each) using a mere $\mathcal{O}(j\log_2^2 j + j n_{qb})$ $\mathtt{CNOT}$ and single-qubit gates, and $3j+\lceil\log_2j\rceil+\mathcal{O}(\max(n_{qb},\log_2 j))$ clean auxiliary qubits. 
The (anti)symmetrizing operator constructed for this purpose depends on the quantum numbers of the additional particle being introduced or ``created''. 
The algorithm improves upon the construction presented in Appendix A of \cite{Galvez-Viruet:2024hry} by almost two powers of $j$ and is on par with the most efficient methods to date, while ideally suited for dynamical memory management with variable particle number. 
We exemplify its applicability with a small selection of problems from hadron physics which are expected to be addressable with upcoming quantum computers.
}

%\pacs{ya veremos}
%%%%%%%%%%%%%%%%%%%%%%%%%%%%%%%%%%%%%%%%%%%%%%%%%%%

\maketitle
\newpage 
%%%%%%%%%%%%%%%%%%%%%%%%
\section{Introduction} \label{sec:intro}
%%%%%%%%%%%%%%%%%%%%%%%%
\subsection{Statement of the problem}
%%%%%%%%%%%%%%%%%%%%%%%%
A cornerstone of Quantum Mechanics is the symmetrization / antisymmetrization of wavefunctions upon exchanging integer / half-integer  quanta.  This is crucial in the atomic {\it Aufbau}, the nuclear shell model, the quark model, and any field theory lifted from them such as Quantum Chromodynamics, to properly count physical states and their energy spectrum. 

With the dawn of the quantum-computer era approaching, 
the simulation of quantum systems is likely to be revolutionized. 
But in multiparticle and field theory simulations, the high expenditure of fully (anti)symmetrizing $j$ identical quanta, with $j!$ possible permutations, can easily run out of control and needs to be tamed by appropriate algorithms.

Part of us~\cite{Galvez-Viruet:2024hry} have developed an (anti)symmetrization algorithm which we have applied to interesting physics problems in hadron physics which would much profit from a full-scale quantum computer (see Section~\ref{sec:problems} for some examples). Because 
main parts of
this algorithm seem to have been rediscovered by further researchers~\cite{Rule:2025ldr}, and because in deploying it we have seen certain improvements which can be effected, we here wish to reappraise it.

In a quantum-register representation, one divides the quantum memory into $n$ registers of size $n_{qb}$ each.  At any time in the computation, $j-1$ of them will contain one of $\tau$ possible orthogonal quantum states or ``modes'' $\alpha_1,\alpha_2\dots \alpha_\tau$. The rest are empty registers, storing a default state $\ket{0^{\otimes n_{qb}}}$,  which we denote as $\ket{\Omega}$. 
(In quark-hadron physics each occupied register can be thought of as a quark or a gluon, in nuclear physics as a nucleon, etc.) 

If each of the modes has occupation number $n_\mu$ (0 or 1 for fermions but any number of bosons), where $\mu \in \{1, \ldots, \tau \}$, the total number of occupied registers is  $j-1=\sum_{\mu}  n_\mu$. The fully symmetric state of the $j-1$ occupied registers is then 
\begin{equation}
\ket{(\alpha_{j-1},\cdots,\alpha_{1})_S} \equiv \frac{1}{\sqrt{(j-1)!\prod_{\mu}n_\mu!}}\sum_{\sigma\in\pi[j-1]} \ket{\alpha_{\sigma(j-1)}}\otimes\cdots\otimes\ket{\alpha_{\sigma(1)}},
\label{symmetricstate}
\end{equation}
where $\pi[j-1]$ is the set of permutations of $j-1$ indices. The objective of the symmetrization algorithm is the implementation of  an operator $\mathtt{S}^{\dagger}_j(\beta)$ which, provided the $j$-th register stores the mode $\beta$ and the previous $j-1$ registers are already symmetric, extends the symmetry to the newly written mode $\beta$
\begin{equation}
\mathtt{S}^{\dagger}_j(\beta)\, \ket{\beta}_j\otimes\ket{(\Phi_{j-1})_S} =\ket{(\beta, \Phi_{j-1})_S}
\label{def:unitary-S}
\end{equation}
where $\ket{(\Phi_{j-1})_S}$ is a symmetric state of $j-1$ occupied registers. Thus (note the inverse order in the productorial),
\begin{equation}
 \ket{(\alpha_j,\cdots,\alpha_{1})_S} = \prod^1_{k=j}\mathtt{S}^{\dagger}_k(\alpha_k)\,\ket{\alpha_j,\alpha_{j-1},\cdots,\alpha_{1}}.
 \label{eq:prepared_state}
\end{equation}
Similar definitions can be written for step-antisymmetrizers $\mathtt{A}^{\dagger}_j(\beta)$, which extend the antisymmetry from $j-1$ to $j$ occupied registers. As particle repetition makes their implementation substantially harder, we dedicate much of the article to the development of step-symmetrizers $\mathtt{S}^{\dagger}_j(\beta)$, and for the antisymmetrizers only quote substantial differences when needed.  The bulk of this article (Sections~\ref{apsec:SDDandS} and~\ref{sec:illustration}) are dedicated to its detailed construction.

%%%%%%%%%%%%%%%%%%%%%%%%%%%%
% Adrian's subsection
\subsection{Extant (anti)symmetrization  algorithms}
%%%%%%%%%%%%%%%%%%%%%%%%%%%%

Quantum algorithms for imposing exchange symmetry in first-quantized encodings differ in their admissible inputs, treatment of repeated occupations, and resource objectives.  Several first-quantized state-preparation algorithms construct an antisymmetric state from a strictly ordered list of occupied modes with $r_1<\cdots<r_n$ in an $M$-dimensional single-particle space through the transformation
\begin{equation}
    \ket{r_1,\dots,r_n} \longmapsto \ket{\Psi_-(r)}=\frac{1}{\sqrt{n!}}\sum_{\pi\in S_n}\mathrm{sgn}(\pi)\ket{r_{\pi(1)},\dots,r_{\pi(n)}}  \ .
\end{equation}
 Restricting the input to ordered configurations ensures that distinct inputs produce orthogonal output states, allowing a unitary implementation on the specified input subspace. An early efficient implementation  prepares a superposition of permutation labels and uses reversible sorting and unsorting  to rearrange the particle registers, with the fermionic sign determined by permutation parity. Its reported complexity is $\mathcal{O}(n^2\log^2{M})$ elementary operations \cite{Abrams:1997gk}. This construction shows that preparing a superposition of $n!$ configurations does not require factorial computational resources (unlike Full Configuration Interaction methods in Quantum Chemistry or Nuclear Physics). A subsequent state-preparation framework also incorporates conversion from second- to first-quantized representations \cite{Ward:2009woa}.

A later sorting-based construction generates permutation records by reversibly sorting auxiliary random labels and then applies the corresponding permutations to the particle registers \cite{Berry:2018ggo}. Any ``collisions'' among the auxiliary labels are detected and removed in a post-selection step, while permutation parity supplies the fermionic sign. Asymptotically optimal sorting networks use $\mathcal{O}(n\log{n})$ comparators, whereas odd-even mergesort networks require $\mathcal{O}(n\log^2{n})$, giving $\mathcal{O}(n\log^2{n}\log{M})$ gate complexity.
As in \cite{Abrams:1997gk},  the algorithm coherently acts  on superpositions of strictly ordered configurations. 
To  work with bosons, however, it is not sufficient to omit the permutation signs, however, since repeated labels (modes with occupancy $n(q)>1$) may appear.

The necessary extension is explicitly addressed by algorithms that allow non-strictly increasing input lists \cite{Liu:2024taj}. The normalized bosonic goal state has the form given in  Eq.~(\ref{symmetricstate}).
When labels are repeated, different permutations can produce the same particle configuration while leaving distinguishable records in the auxiliary registers. These redundant records must therefore be coherently erased.

 Starting from ordered list with possible repetitions, the resulting algorithms achieve $\mathcal{O}(\log{n}\log\log n)$ depth with $\mathcal{O}(n\log{n}+\log{M})$ ancillary qubits for a single classical input list, with high success probability, and $\mathcal{O}(\log^3{n}\log\log{n})$ depth with $\mathcal{O}(n\log{n})$ ancillaries for arbitrary coherent superpositions, with unit success probability \cite{Liu:2024taj}.
 
To compare these reported counts with our new algorithm is difficult. We have to take into account that the depth of those quantum circuits, defined as the number of consecutive layers of gates which cannot be parallelized, is generally smaller than their total gate count. Therefore, the depth estimates reported for them in the literature should not be directly compared with the gate complexity of the iterative symmetrization algorithm presented here. Unfortunately, a more detailed gate-level decomposition of the algorithm in~\cite{Liu:2024taj} would be necessary to establish a meaningful comparison
\footnote{ Those quoted depth bounds use asymptotically optimal sorting networks and suppress the additional comparator-depth dependence on register width. Replacing these networks by bitonic or odd-even mergesort networks adds a logarithmic factor to the sorting depth.}. From cursory examination, we think it is safe to state that our algorithm is at 
least as efficient as the best known  methods.

A complementary approach builds exchange symmetry recursively, incorporating one particle at a time into an already symmetric or antisymmetric state. This is the crux of our earlier algorithm~\cite{Galvez-Viruet:2024hry}.  We introduced step-symmetrizers and step-antisymmetrizers as part of a particle-register implementation of creation and annihilation operators, and described unitary implementations of these transformations \cite{Galvez-Viruet:2024hry}. In particular, the step-antisymmetrizer has the form
\begin{equation}
    \mathcal{A}_{k}=\frac{1}{\sqrt{k}}\left(I-\sum_{j=1}^{k-1}\mathcal{P}_{jk}\right) \ ,
    \label{eq:s-antisymmetrizer}
\end{equation}
where $\mathcal{P}_{jk}$ swaps particle registers $j$ and $k$,  and the added orbital is orthogonal to those previously occupied (think, for example, of momentum modes).

A recently reported recursive construction
\cite{Rule:2025ldr} 
assumes access to preparation circuits $U_j\ket{0}^{\otimes q}=\ket{\phi_j}$, with $q=\lceil\log_2 M\rceil$, for mutually orthogonal orbitals. Its auxiliary registers are subsequently uncomputed using the inverse preparation circuits $U_j^\dagger$ and orbital orthogonality. No numerical ordering of the orbitals is required. For computational-basis orbitals, the reported $T$-gate count scales as $\mathcal{O}(n^2\log{M})$. For nontrivial orbitals, the preparation adopted in \cite{Rule:2025ldr} gives $\mathcal{O}(n^2\sqrt{M})$ scaling, with an additional $\mathcal{O}(\sqrt{M})$ dirty ancillaries. A measurement-and-feed-forward variant retains unit success probability and approximately halves, on average, the number of orbital-preparation and inverse-preparation operations that dominate the cost in this regime.

These approaches therefore address different versions of the state-preparation problem. Sorting-based methods provide favorable particle-number scaling for ordered configurations, general bosonic symmetrization requires explicit treatment of repeated occupations, and recursive fermionic constructions incorporate independently specified orbital-preparation circuits. 

We are now ready to present the novel improvement which we here propose to our earlier setup. In summary, we present a novel {\tt UDD} algorithm which improves our earlier implementation by almost a factor of $j^2$ by using the Shukla-Vedula algorithm~\cite{Shukla:2023xla} to prepare uniform superpositions, and improving the final steps in appendix A.2.a of \cite{Galvez-Viruet:2024hry} to prepare register $C$.

%%%%%%%%%%%%%%%%%%%%%%%%%%%%%%
 \section{Modification of our earlier step-symmetrization algorithm} \label{apsec:SDDandS}
%%%%%%%%%%%%%%%%%%%%%%%%%%%%%%

In our earlier algorithm~\cite{Galvez-Viruet:2024hry}, the implementation of the (unitary) step symmetrizer $\mathtt{S}^\dagger_j(\alpha)$  unnecessarily increased the asymptotic costs for large particle number, mostly due to the use of the $\mathtt{LL}$ (``Locate the Largest'') algorithm and the expensive $\mathtt{UDD}$ subroutine to which we dedicate the next subsection~\ref{subsec:UDD}. We here revisit its implementation as well as that of the step (anti)symmetrizers, in subsection~\ref{subsec:step}, to later demonstrate a gain in  asymptotic costs.
%%%%%%%%%%%%%%%%%%%%%%%%%%%%%
\subsection{UDD subroutine}
\label{subsec:UDD}
%%%%%%%%%%%%%%%%%%%%%%%%%%%%%
The  {\tt UDD} ``uniform digit decomposition'' algorithm ({\tt SDD} in \cite{Galvez-Viruet:2024hry}) is a key subroutine in the implementation of the step symmetrizes that disentangles the auxiliary registers needed to account for mode multiplicities greater than 1. It consists on the decomposition of a binary number $b$ of Hamming weight $k$ (number of nonvanishing bits $k=\sum_i b_i$)~\cite{Hamming:1950} into its binary digits $b_i$ in superposition. The result of the decomposition is stored in an auxiliary register $C$ initially in the reference state $\ket{0}_C$,
\begin{equation}
\mathtt{UDD}\ket{0}_C\ket{b}_B = \begin{cases}\frac{1}{\sqrt{k}}\Big(\sum^{n}_{i=1}b_i\ket{ 2^{i-1}}_C\Big)|b\rangle_B, & k >0,\\
\ket{0}_C\ket{b}_B& k= 0,
\end{cases}
\end{equation}
where $b=\sum^{n}_{i=1} b_i 2^{i-1}$ has $n$ digits. The state $\ket{2^{i-1}}$ should be understood as having qubit $i$ (counting from right to left and starting at $1$) in state $\ket{1}$ and all the others in state $\ket{0}$. Denoting by $l$ the set of indices $i$ for which $b_i = 1$ and denoting $l_j$ the $j$th of such indices, i.e. $l=\{i|b_i=1\}=\{l_1,l_2,...,l_k\}$ we have
\begin{equation}\label{delsaLs}
\Big(\sum^{n}_{i=1}b_i\ket{ 2^{i-1}}_C\Big)\otimes \Big|\sum^{n}_{j=1} b_j 2^{j-1} \Big\rangle_B = \Big(\sum^{k}_{i=1}\ \Big|2^{l_i-1}\Big\rangle_C\Big)\otimes \Big|\sum^{k}_{j=1}2^{l_j-1} \Big\rangle_B .
\end{equation}
(In $|\ \rangle_C$ the sum runs over kets, not over bits inside the same ket as in $|\ \rangle_B$.)

The algorithm is explained by providing the evolution of an ancillary state through several steps and substeps $\ket{\psi_1}\to \ket{\psi_2}\to \dots \ket{\psi_{6,3}}$. We will omit $\ket{\ }_B$ which remains unentangled, until step 6. These instructions should be programmed in the order given:

\begin{enumerate}
\item The decomposition register $C$ has $n$ qubits. We introduce further auxiliary registers $E$ with $\text{max}(n-1,\lceil\log_2(n+1)\rceil)$ qubits and $D$ with $\lceil\log_2 n\rceil$ qubits. Throughout the algorithm, $C_i$ denotes the $i$th qubit of register $C$ (and likewise for the other registers), counting from right to left and starting at $1$. A ket containing all those ancillary registers is
\begin{equation}
\ket{\psi_1} = \ket{0}_E\ket{0}_D\ket{0}_C.
\end{equation}
\medskip 

\item Calculate $k=\sum_i b_i$ and store it in register $E$,
\begin{equation}
\ket{\psi_2} = \ket{k}_E\ket{0}_D\ket{0}_C.
\end{equation}
Note that a binary number of length $n$ contains at most $n$ ones. Since the number of ones can range from $0$ to $n$, there are $n+1$ possible values of $k$. This step would only require register $E$ to have $\lceil\log_2(n+1)\rceil$ qubits. The reason it instead allocated $n-1$ qubits is for later usage in step 6. For  $n=2$, we have that $n-1<\lceil\log_2(n+1)\rceil$, so only in this particular case, register $E$ must have $\lceil\log_23\rceil=2$ qubits and not one.

\textbf{Costs: } This preparation can be achieved with $\cO(n)$ unit adders on $\cO(\log_2 n)$ qubits of $E$, which gives a total cost of $\cO(n\log_2 n)$ single-qubit and $\mathtt{CNOT}$ gates and $\cO(\log_2(n)
)$ extra auxiliary qubits. 
\medskip

\item  Unless $k=0$, apply the following transformation to register $D$,
\begin{equation}
\ket{\psi_2} = \ket{k}_E\ket{0}_D\ket{0}_C \rightarrow \ket{\psi_3}=\ket{k}_E\Big(\frac{1}{\sqrt{k}}\sum^{k-1}_{i= 0}\ket{i}_D\Big)\ket{0}_C,
\end{equation}
to do this, use the Shukla-Vedula algorithm~\cite{Shukla:2023xla} to prepare uniform superpositions (substituting the generic $M$ of that algorithm by the variable $k$ here needed). 

The algorithm starts with the  value of $k$ encoded in register $|\ \rangle_E$ and as $k$ is unknown outside, unlike in the original Shukla-Vedula formulation~\cite{Shukla:2023xla}, one proceeds with a series of single-controlled $\mathtt{R}_Y$ rotations, Hadamard gates and $\mathtt{CNOT}$s. Its reported asymptotic cost is of order $\cO(\log_2 k)$ gates. In our case, the  values of $k$ different from zero are stored in a quantum register of $\log_2 n$ qubits, therefore their single-controls should be transformed into multi-controlled gates over $\log_2 n + 1$ qubits, with an asymptotic cost of $\cO(\log_2 n \log_2 k)$ gates and  $\cO(\log_2 n)$ auxiliary qubits.  

The total {\bf cost},  since there are $n$ possible values of $k>0$, is $\cO(n\log^2_2 n)$ gates and $\cO(\log_2 n)$ auxiliary qubits.
\medskip

\item Loop over $i$ from $i = 1$ to $i = n$,
\begin{enumerate}
\item[$$4.1$$] If register $D$ is $\ket{0}_D=\ket{0\cdots 0}$ and qubit $B_i$ is in state $\ket{1}$, apply an $\mathtt{X}$ gate to the corresponding qubit $C_i$.
Such an $\mathtt{X}$ gate, controlled by $\log_2 n$ qubits, has a {\bf cost} of $\cO(\log_2 n)$ single-qubit and $\mathtt{CNOT}$ gates.
\smallskip

\item[$$4.2$$] If $b_i = 1$ decrease  register $D$ by $1$ modulo $2^{\lceil\log_2n\rceil}=2^{n_D}$. Note that a unit decrease of $\ket{0}_D$ then gives $\ket{2^{n_D}-1}_D$, and therefore it is necessary to perform $2^{n_D}$ unit decreases to go a complete cycle and return to a given initial value.\\
This controlled decrease, operating on the $\cO(\log_2 n)$ qubits of $|\ \rangle_D$, then entails 
\textbf{costs} of $\cO(\log_2 n)$ single-qubit and $\mathtt{CNOT}$ gates .  
\end{enumerate}
End the loop by adding $\ket{k}_E$ modulo $2^{\lceil\log_2 n\rceil}$ to register $D$. The result is
\begin{equation}\label{paso4}
\ket{\psi_4} = \ket{k}_E\frac{1}{\sqrt{k}}\Big(\sum^{k}_{j=1}\ket{j-1}_D\ket{ 2^{l_j-1}}_C\Big).
\end{equation}
The remaining steps will basically handle the ``uncomputation'' (reversing of the operations effected) of registers $E$ and $D$ to get the clean result on $C$. 

$\textbf{Costs: }$ The addition on registers of size $\cO(\log_2 n)$ qubits require $\cO(\log_2 n)$ single-qubit and $\mathtt{CNOT}$ gates.
\medskip

\item Uncompute register $E$. This takes $\ket{k}_E$ in Eq.~(\ref{paso4}) to a $\ket{\psi_5}$ in which it is set to $\ket{0}_E$, free for recycling, with no other change.
\medskip

\item The following three steps disentangle registers $D$ from $C$. Register $C$ is in a superposition of states with only one qubit in state $\ket{1}$. This qubit coincides with a qubit $B_m=\ket{1}$ such that the number of qubits $B_i=\ket{1}$ with $i<m$ is equal to the state of $D$. For each $i$ from $1$ to $n-1$, the first step initiates qubits $E_1$ to $E_i$ if qubit $C_{i+1}=\ket{1}$, the second step then compares qubits $E_j$ and $B_j$, with $j$ from $1$ to $n-1$ and for each coincidence diminishes the value of $D$ by $1$. A final step disentangles register $E$:
\smallskip

\begin{enumerate}
\item[$$6.1$$] 
First, if $C_{n} = \ket{1}$ initiate qubit $E_{n-1}$ to $\ket{1}$. 

\noindent Now, iterate from $i=1$ to $i = n-2$. If $E_{n-i} = \ket{1}$ or the qubit $C_{n-i} = \ket{1}$, then also initiate qubit $E_{n-1-i}$. This gives,
\begin{equation}
\ket{\psi_{6,1}} = \frac{1}{\sqrt{k}}\Big(\sum^{k}_{i=1}\Big|2^{l_i-1}-1\Big\rangle_E\Big|i-1\Big\rangle_D\Big| 2^{l_i-1}\Big\rangle_C\Big)\Big|\sum^{k}_{j=1} 2^{l_j-1} \Big\rangle_B.
\end{equation}
\textbf{Costs: } $\cO(n)$ $\mathtt{OR}$ tests over a constant number of qubits cost $\cO(n)$ single qubit gates and $\mathtt{CNOT}$.
\smallskip

\item[$$6.2$$] For each $m$ from $1$ to $n$ such that $C_m=\ket{1}$, the state of $D$ is equal to how many qubits $B_i$ are $1$ with $i<m$, furthermore, register $E$ has qubits $E_1$ to $E_{m-1}$ equal to $\ket{1}$. We can therefore disentangle $D$ by subtracting $1$ each time two qubits of $B$ and $E$ with the same index are both $\ket{1}$,
\begin{equation}
\ket{\psi_{6,2}} = \frac{1}{\sqrt{k}}\Big(\sum^{k}_{i=1}\Big|2^{l_i-1}-1\Big\rangle_E\ket{0}_D\Big| 2^{l_i-1}\Big\rangle_C\Big)\Big|\sum^{k}_{j=1} 2^{l_j-1} \Big\rangle_B,
\end{equation}
\textbf{Costs: } $\cO(n)$ subtraction circuits on $\cO(\log_2 n)$ qubits controlled over a constant number of qubits cost $\cO(n\log_2n)$ single-qubit gates and $\mathtt{CNOT}$ and $\cO(\log_2 n)$ extra auxiliary qubits.
\smallskip

\item[$$6.3$$] Finally, repeating $6.1$ in reverse to disentangle $E$ gives,
\begin{equation}
\ket{\psi_{6,3}} = |0\rangle_E\ket{0}_D\frac{1}{\sqrt{k}}\Big(\sum^{k}_{i=1}\Big| 2^{l_i-1}\Big\rangle_C\Big)|\sum^{k}_{j=1} 2^{l_j-1} \Big\rangle_B\ .
\end{equation}
Registers $D$ and $E$ are now disentangled and can be ignored; the state of $|\ \rangle_C\otimes |\ \rangle_B$ is exactly the right hand side of Eq.~(\ref{delsaLs}) which then closes the algorithm, its validity and efficacy having been demonstrated.
\end{enumerate}
\end{enumerate}
\textbf{Total costs: } Examining  all steps, we find that the algorithm requires $\cO(n\log^2_2 n)$ gates and $j+\lceil\log_2j\rceil+\cO(\log_2 n)$ auxiliary qubits, a substantial improvement with respect to the $\cO(n^3\log n)$ complexity reported in \cite{Galvez-Viruet:2024hry}.

%%%%%%%%%%%%%%%%%%%%%%%%%%%%%%%
\subsection{Step-(anti)symmetrizer implementation}
\label{subsec:step}
%%%%%%%%%%%%%%%%%%%%%%%%%%%%%%%

We now revisit the implementation of the step symmetrizer already discussed in appendix A.2.b of~\cite{Galvez-Viruet:2024hry} providing a more detailed proof of the asymptotic costs. Substantial differences with respect to the implementation of step-antisymmetrizers are pointed out. 

We now incorporate the main particle register to the auxiliary ones discussed in the previous Subsection~\ref{subsec:UDD}.
Proceeding in alphabetic order, the input particle registers are collectively denoted by $A$, and we refer to the $i$th one as subregister $A_i$. In addition to $A$, we introduce two auxiliary qubit registers, $B$ (which will keep track of repeatedly occupied boson modes) and $C$, each consisting of $j$ qubits, where $j$ is the number of occupied registers.\footnote{In general, the state is a superposition of components with different numbers of occupied registers, and the full algorithm  which we have implemented controls on the number of active registers. We omit these details here for conciseness.} 
The implementation of $\mathtt{S}^{\dagger}_j(\alpha)$ 
of Eq.~(\ref{def:unitary-S})
proceeds as follows.
\begin{enumerate}
\item The first step is then to initialize register $B$ in the state $|0\rangle_B$ and, provided mode $\alpha$ lies in the last occupied subregister $A_j$, initialize register $C$ in the state $\frac{1}{\sqrt{j}}\sum_{k=1}^{j} |2^{k-1}\rangle_C$ (this is so there is exactly a 1 in each term of the linear superposition, with the rest of the qubits being set to zero). If this is so,
\begin{equation}
\label{Psi1aux}
\ket{\phi_1} = \frac{1}{\sqrt{j}}\sum^{j}_{k=1}|2^{k-1}\rangle_C\otimes |0\rangle_B\otimes\ket{\alpha, (\Phi_{j-1})_S}_A.
\end{equation}

(In the antisymmetrization case, each particle configuration must be unique, and therefore mode $\alpha$ exclusively sits on register $j$ by hypothesis\footnote{Such hypothesis can be lifted at the cost of adding further controls in the preparation of register $C$.}. This renders register $B$ unnecessary. The preparation of register $C$ proceeds in the same way.)

To execute the transformation of Eq.~(\ref{Psi1aux}) we apply $\cO(j)$ single-qubit gates controlled by the $\cO(n_{qb})$ qubits of a single subregister of $A$, which entails $\textbf{costs}$ of $\cO(j n_{qb})$ {\tt CNOT} and single-qubit gates and $\cO(n_{qb})$ auxiliary registers.
\item  This step is only necessary for the symmetrization case, and absent for fermions. Register $A$ consists of $j$ sub-registers of size $n_{qb}$ storing different particle modes.  We use an auxiliary register $B$ to mark the sub-registers $A_i$ which contain mode $\alpha$, so ($A_i=\ket{\alpha}$ will correspond to $B_i=\ket{1}$ whereas $A_i\neq\ket{\alpha}$ to $B_i=\ket{0}$). In this way, instead of having to work with $j$ sub-registers of $n_{qb}$ qubits, we work with a single register $B$ of $j$ qubits. This is done by applying one multi-controlled gate $\mathtt{C}^{n_{qb}}\mathtt{X}(\alpha)$ controlled on each subregister $A_i$, and with target qubit $B_i$, for $i=1$ to $i=j$; see figure~\ref{fig:paso2aplicacionS}.

%To be consistent with Eq.~(\ref{def:unitary-S}),the initialization of register $B$ should be controlled on mode $\alpha$ being stored in subregister $j$ of A. For doing so more efficiently, we first apply a multi-controlled gate $\mathtt{C}^{n_{qb}}\mathtt{X}(\alpha)$ to initiate qubit $B_j$, see first gate of figure~\ref{fig:paso2aplicacionS}. Next, we store the remaining positions of $\alpha$ by applying one multi-controlled gate $\mathtt{C}^{n_{qb}+1}\mathtt{X}(\alpha)$ controlled on each subregister $A_i$ and qubit $B_j$, and with target $B_i$, for $i=1$ to $i=j-1$; see remaining gates of figure~\ref{fig:paso2aplicacionS}. From now on, we omit register $A$ as the information we needed has been transferred to the auxiliary register $B$. 

Provided that previously occupied particle registers are maintained to be fully symmetrized at all times, and that mode $\alpha$ sits on register $j$, the auxiliary registers will be transformed into (up to normalization),
\begin{equation}
\ket{\phi_2}_{\text{aux}} =  \sum^{j}_{k=1}|2^{k-1}\rangle_C\otimes \sum_{\substack{1\leq l_1<l_2 <  \\ ... < l_{n_1}\leq j-1}}\big|(j,l_{n_1},...,l_2,l_1)\big\rangle_B,
\end{equation}
the notation means that qubits $B_{l_1},\cdots B_{l_{n_1}},B_{j}$ have rotated to state $\ket{1}$, with the remaining, omitted indices indicate qubits which remain unchanged. $n_1$ is the multiplicity of mode $\alpha$ for register $1$ to $j-1$.

$\textbf{Costs:}$ $\cO(j n_{qb})$ single-qubit gates and $\cO(n_{qb})$ auxiliary qubits.

\begin{figure}[h]
    \centering
    \includegraphics[width=0.8\linewidth]{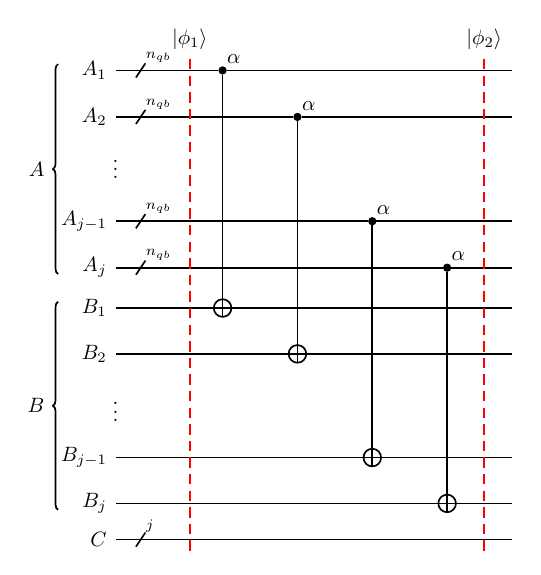}
    \caption{Multicontrolled operation to store the positions of the $\alpha$ mode in register $B$. Qubit $B_i$ is flipped to $\ket{1}$ when sub-register $A_i$ stores mode $\ket{\alpha}$. Notation: the $\bullet^\alpha$ denotes conditioning on the multi-qubits state $\ket{\alpha}$ rather than in a single qubit state $\ket{1}$.}% represented by $\bullet$.}
    \label{fig:paso2aplicacionS}
\end{figure}

\item Symmetrize the $j$-th register with all the others by applying the operator 
\begin{equation} 
U_{\rm perm}\equiv \sum^{j}_{k=1}|2^{k-1}\rangle\langle 2^{k-1} |_C \otimes P^{B}_{j,k} \otimes P^{A}_{j,k} + \Big(I^C-\sum^{j}_{k=1}|2^{k-1}\rangle\langle 2^{k-1} |_C\Big)\otimes I^B\otimes I^A\ ,
\label{eq:defUperm}
\end{equation}
where the permutator $P^{B}_{l,k}$ interchanges qubits $l$ and $k$ of register $B$ and $P^{A}_{l,k}$ interchanges particle registers $l$ and $k$ of $A$. This construction guarantees unitarity. After its application, the auxiliary registers are (we drop normalization constants and the superindex $B$ from the permutators),
\begin{equation}
|\phi_3\rangle_\text{aux} = \sum^{j}_{k=1}|2^{k-1}\rangle_C \otimes \sum_{\substack{1\leq l_1<l_2 <  \\ ... < l_{n_1}\leq j-1}} P_{j,k}\big|(j,l_{n_1},...,l_2,l_1)\big\rangle_B.
\label{eq:4.2}
\end{equation}
If $k = l_i$, $i\in(1,...,n_1)$ or $k=j$, the permutator interchanges two qubits of $B$ whose states are $\ket{1}$  (and two particle registers with mode $\alpha$) and no change is produced. Therefore Eq.~(\ref{eq:4.2}) can be rewritten as:
\begin{align}
|\phi_3\rangle_\text{aux} & = \sum_{\substack{1\leq l_1<l_2 <  \\ ... < l_{n_1}\leq j-1}} \Bigg[\bigg(\sum^{n_1}_{m=1}|2^{l_m-1}\rangle_C +|2^{j-1}\rangle_C\bigg)\otimes\big|(j,l_{n_1},...,l_2,l_1)\big\rangle_B \nonumber \\
& + \bigg(\sum^{j-1}_{\substack{k=1, \\ k\notin(l_1,..,l_{n_1})}}|2^{k-1}\rangle_C \otimes P_{j,k}\big|(j,l_{n_1},...,l_2,l_1)\big\rangle_B\bigg)\Bigg],
\label{eq:4.3}
\end{align}
and the second line encompasses the remaining terms. Focusing on those, the interchange entails a new  qubit in state $1$ from positions $1$ to $j-1$ (and a $0$ in position $j$). For the case in which $k\in(1,\cdots, l_1-1)$: 
\begin{align}
\sum_{\substack{1\leq l_1<l_2 <  \\ \cdots < l_{n_1}\leq j-1}} \sum^{j-1}_{\substack{k=1, \\ k\notin(l_1,..,l_{n_1})}}(\cdots) & = \sum_{\substack{1\leq l_1<l_2 <\cdots  \\  < l_{n_1}\leq j-1}}\sum^{j-1}_{\substack{k=l_1+1, \\ k\notin(l_2,..,l_{n_1})}} (\cdots) + \sum_{\substack{1 < l_1<l_2 <\cdots  \\  < l_{n_1}\leq j-1}}\sum^{l_1-1}_{k=1}(\cdots), 
\label{eq:4.4}
\end{align} 
applying the change $k\rightarrow l_1$, $l_i\rightarrow l_{i+1}$ to the last two summands of Eq.~(\ref{eq:4.4}), 
\begin{equation}
\sum_{\substack{1\leq l_1<l_2 <  \\ \cdots < l_{n_1}\leq j-1}} \sum^{j-1}_{\substack{k=1, \\ k\notin(l_1,..,l_{n_1})}}(\cdots) = \sum_{\substack{1\leq l_1<l_2 <\cdots  \\  < l_{n_1}\leq j-1}}\sum^{j-1}_{\substack{k=l_1+1, \\ k\notin(l_2,..,l_{n_1})}}(\cdots) + \sum_{\substack{1 \leq l_1 < l_2 <\cdots  \\ <l_{n_1+1}\leq j-1}}(\cdots), 
\end{equation} 
and the second line of Eq.~(\ref{eq:4.3}) now reads,
\begin{align}
\sum_{\substack{1\leq l_1<l_2 <  \\ \cdots < l_{n_1}\leq j-1}}&  \sum^{j-1}_{\substack{k=1, \\ k\notin(l_1,..,l_{n_1})}} |2^{k-1}\rangle_C \otimes P_{j,k}\big|(j,l_{n_1},...,l_2,l_1)\big\rangle_B\nonumber \\
=&\sum_{\substack{1\leq l_1<l_2 <\cdots  \\  < l_{n_1}\leq j-1}}\sum^{j-1}_{\substack{k=l_1+1, \\ k\notin(l_2,..,l_{n_1})}}|2^{k-1}\rangle_C \otimes P_{j,k}\big|(j,l_{n_1},...,l_2,l_1)\big\rangle_B\nonumber \\
+&\sum_{\substack{1 \leq l_1 < l_2 <\cdots  \\ <l_{n_1+1}\leq j-1}}|2^{l_1-1}\rangle_C \otimes \big|(l_{n_1+1},...,l_2,l_1)\big\rangle_B.
\end{align}
Repeating the same transformation for each of the $n_1$ indices we have
\begin{align}
& \sum_{\substack{1\leq l_1<l_2 <  \\ \cdots < l_{n_1}\leq j-1}}\bigg(\sum^{j-1}_{\substack{k=1, \\ k\notin(l_1,..,l_{n_1})}}|2^{k-1}\rangle_C \otimes P_{j,k}\big|(j,l_{n_1},...,l_2,l_1)\big\rangle_B\bigg) \nonumber \\
& = \sum_{\substack{1\leq l_1<l_2 <  \\ \cdots < l_{n_1+1}\leq j-1}}\bigg(\sum^{n_1+1}_{m=1}|2^{l_m-1}\rangle_C\bigg)\otimes\big|(l_{n_1+1},l_{n_1},...,l_2,l_1)\big\rangle_B,
\end{align}
so that 
\begin{align}
|\phi_3\rangle_\text{aux} & = \sum_{\substack{1\leq l_1<l_2 <  \\ ... < l_{n_1}\leq j-1}}\bigg(\sum^{n_1}_{m=1}|2^{l_m-1}\rangle_C +|2^{j-1}\rangle_C\bigg)\otimes\big|(j,l_{n_1},...,l_2,l_1)\big\rangle_B \nonumber \\
& +  \sum_{\substack{1\leq l_1<l_2 <  \\ \cdots < l_{n_1+1}\leq j-1}}\bigg(\sum^{n_1+1}_{m=1}|2^{l_m-1}\rangle_C\bigg)\otimes\big|(l_{n_1+1},l_{n_1},...,l_2,l_1)\big\rangle_B. 
\label{eq:4.8}
\end{align}
Now, realizing that the first line of $|\phi_3\rangle_\text{aux}$ corresponds to the case in which $l_{n_1+1} = j$ we have,
\begin{equation}
|\phi_3\rangle_\text{aux} =  \sum_{\substack{1\leq l_1<l_2 <  \\ \cdots < l_{n_1+1}\leq j}}\bigg(\sum^{n_1+1}_{m=1}|2^{l_{m}-1}\rangle_C\bigg)\otimes\big|(l_{n_1+1},l_{n_1},...,l_2,l_1)\big\rangle_B.
\end{equation}
The permutations of the $n_1+1$ indices ensure that the particle registers are fully symmetrized, so all that remains is to disentangle the auxiliary registers. Note that no further controls are needed to prevent unwanted executions when mode $\alpha$ does not lie in register $A_j$, since in this case registers $B$ and $C$ wouldn't be initialized.

In the anti-symmetrization case, the first term in the operator Eq.~\eqref{eq:defUperm} should be changed to include a $-$ sign whenever $k<j$ and there is no need for the permutator over register $B$. At the end of this step, the content of register $j$ will be interchanged for the content of all the other registers, thus achieving antisymmetrization. The last step is then to disentangle register $C$.

\textbf{Costs: } Same for symmetrizers and antisymmetrizers, register $C$ has $j$ qubits, each of which is used as control over two-particle register $\mathtt{SWAP}$ gates that costs $\cO(n_{qb})$ Toffoli gates and $\cO(n_{qb})$ extra auxiliary qubits. Since each Toffoli can be decomposed into a constant number of single-qubit gates and $\mathtt{CNOT}$s, the total {\bf cost} is $\cO(j n_{qb})$ single-qubit and $\mathtt{CNOT}$ gates and $\cO(n_{qb})$ extra auxiliary qubits. 

\item This step is only needed for the symmetrization case. Since registers $B$ and $C$ are exactly of the form generated by the $\mathtt{UDD}$ algorithm, that is
\begin{equation}
|\phi_3\rangle_\text{aux} =  \sum_{\substack{1\leq l_1<l_2 <  \\ \cdots < l_{n_1+1}\leq j}}\bigg(\sum^{n_1+1}_{m=1}|2^{l_{m}-1}\rangle_C\bigg)\otimes\big|\sum^{n_1+1}_{m=1}2^{l_m-1}\big\rangle_B,
\end{equation}
register $C$ can be disentangled by applying the {\tt UDD} algorithm in reverse:
\begin{equation}
|\phi_4\rangle_\text{aux} = \mathtt{UDD}^{-1}|\phi_3\rangle_\text{aux} =|0\rangle_C \otimes\sum_{\substack{1\leq l_1<l_2 <  \\ \cdots < l_{n_1+1}\leq j}}\big|\sum^{n_1+1}_{m=1}2^{l_m-1}\big\rangle_B.
\end{equation}

\textbf{Costs}: $\cO(j\log^2_2 j)$ gates and $j+\lceil\log_2j\rceil+\cO(\log_2 j)$ extra auxiliary qubits.
\item For the (anti)symmetrization case, disentangle register ($C$) $B$ by undoing step 2, that is, for $i=1$ to $i=j$ apply a multi-controlled $\mathtt{C}^{n_{qb}}\mathtt{X}(\alpha)$, with the controls being the $n_{qb}$ qubits of subregister $A_i$ and the target being the ($C_i$) $B_i$ qubit. 

\textbf{Costs: } $\cO(j n_{qb})$ single-qubit gates and $\cO(n_{qb})$ auxiliary qubits.
\item[$$$$] Summing up all costs, the {\bf total asymptotic complexity} for symmetrization is $\cO(j\log^2_2 j + jn_{qb})$ gates and $3j+\lceil\log_2j\rceil+\cO(\text{max}(n_{qb},\log_2 j))$ auxiliary qubits and for antisymmetrization is $\cO(jn_{qb})$ gates and $j+\cO(n_{qb})$ auxiliary qubits.
\end{enumerate}

\newpage
%%%%%%%%%%%%%%%%%%%%%%%%%%%%%%
\section{Worked examples of the step symmetrization and {\tt UDD} algorithms} \label{sec:illustration}

%%%%%%%%%%%%%%%%%%%%%%%%%%%%%%
\subsection{Step-symmetrizers}
%%%%%%%%%%%%%%%%%%%%%%%%%%%%%%
As an example, let us see how to symmetrize a state with three bosons. The goal of the ${\tt S}_3^\dagger(P)$ operator is to construct the symmetric state of three particles. In this example, we consider three registers and two different modes, $P$ and $q$, with the $P$ mode appearing twice. We assume that registers $1$ and $2$ are already symmetrized

\begin{equation}
    |(P,q)_S\rangle=
    \frac{1}{\sqrt{2}}\left(
        |P\rangle_2|q\rangle_1+
        |q\rangle_2|P\rangle_1\right).
\end{equation}
We want to add register $3$, which contains the known mode $P$. Thus, the initial state (stored in register $A$) is
\begin{equation}
    |\phi_{0}\rangle=
    |P\rangle\otimes |(P,q)_S\rangle=
    \frac{1}{\sqrt{2}} \left(
        |PPq\rangle_A+|PqP\rangle_A\right),
\end{equation}
where we use the notation 
$|x_3x_2x_1\rangle\equiv
|x_3\rangle_3|x_2\rangle_2|x_1\rangle_1$. Following  Eq.~(\ref{Psi1aux}), to keep track of all indices we introduce two auxiliary registers $B$ and $C$, each containing three qubits, initialized as

\begin{equation}
    |B\rangle = |000\rangle_B,
    \qquad
    |C\rangle = \frac{1}{\sqrt{3}}\left(
        |100\rangle_C+
        |010\rangle_C+
        |001\rangle_C\right).
\end{equation}
So that
\begin{equation}
    |\phi_1\rangle=\frac{1}{\sqrt{6}}\left(
        |100\rangle_C+
        |010\rangle_C+
        |001\rangle_C\right) |000\rangle_B \left(
        |PPq\rangle_A+|PqP\rangle_A\right).
\end{equation}
Register $B$ is used to mark the positions containing the known mode $P$. For this, a ${\tt C}^{n_{\rm qb}}{\tt X}(P)$ gate, controlled by subregister $A_i$, is applied to each qubit $B_i$. The gate flips $B_i$ whenever $A_i$ contains the mode $P$. This operation can be understood as a $   \tt{CNOT}_{A_i\rightarrow B_i}$. Therefore,

\begin{align}
    |000\rangle_B|PPq\rangle_A&\longrightarrow|110\rangle_B|PPq\rangle_A,
    \\
    |000\rangle_B|PqP\rangle_A&\longrightarrow|101\rangle_B|PqP\rangle_A,
\end{align}
which gives
\begin{equation}
    |\phi_2\rangle =
    \frac{1}{\sqrt{6}}\left(
        |100\rangle_C+
        |010\rangle_C+
        |001\rangle_C\right)
    \left(
        |110\rangle_B|PPq\rangle_A
        +
        |101\rangle_B|PqP\rangle_A\right). \label{eq-step2}
\end{equation}
Now, we apply the operator of Eq.~(\ref{eq:defUperm}) with $j=3$

%Next, we change register $C$ to a superposition over the possible placements of the flipped qubit, 

%\begin{equation}
    %|100\rangle_C
    %\longrightarrow
    %\frac{1}{\sqrt{3}}\left(
        %|100\rangle_C+
        %|010\rangle_C+
        %|001\rangle_C\right), \label{eq-superpositionC}
%\end{equation}
%so that,
%\begin{align}
    %|\phi_2\rangle=\frac{1}{\sqrt{6}}\left(
        %|100\rangle_C+
        %|010\rangle_C+
        %|001\rangle_C\right)\left(
        %|110\rangle_B|PPq\rangle_A+
        %|101\rangle_B|PqP\rangle_A\right).
%\end{align}
\begin{equation}
    U_{\mathrm{perm}}=
    \sum_{k=1}^{3}
    |2^{k-1}\rangle\langle2^{k-1}|_C\otimes P^B_{3k} \otimes P^A_{3k} +  \Big(I^{\otimes 3}-\sum^{3}_{k=1}|2^{k-1}\rangle\langle 2^{k-1} |_C\Big)\otimes I^B\otimes I^A,
\end{equation}
where the permutation  $P^A_{3k}$ exchanges particle registers $3$ and $k$, while $P^B_{3k}$ does the same for the auxiliary register $B$. Specifically, depending on the state of register $C$, the permutation operators applied act as follows,
\begin{align}
    |100\rangle_C  &\longrightarrow P^B_{33}\otimes P^A_{33} =I,
    \\
    |010\rangle_C &\longrightarrow P^B_{32}\otimes P^A_{32},
    \\
    |001\rangle_C  &\longrightarrow P^B_{31}\otimes P^A_{31}.
\end{align}
Therefore, the first component leaves the state invariant, the second exchanges registers $3$ and $2$, and the third exchanges registers $3$ and $1$. Applying these permutations, we obtain
\begin{align}
    |\phi_3\rangle= \frac{1}{\sqrt{3}} \Bigg[&
         \frac{|100\rangle_C+|010\rangle_C}{\sqrt{2}}
        |110\rangle_B|PPq\rangle_A
        \nonumber\\&+
        \frac{|100\rangle_C+|001\rangle_C}{\sqrt{2}}
        |101\rangle_B|PqP\rangle_A
        \nonumber\\ &+
        \frac{|010\rangle_C+|001\rangle_C}{\sqrt{2}}
        |011\rangle_B|qPP\rangle_A\Bigg].
    \label{eq:example_after_permutations}
\end{align}
Equation~\eqref{eq:example_after_permutations} has the structure generated by the {\tt UDD} operator. As an example, subsection \ref{subsubsec:UDD} demonstrates that the {\tt UDD} generates the state corresponding to the first line of Eq.~\eqref{eq:example_after_permutations} from the state $\ket{000}_C\ket{110}_B$, therefore disentangling register $C$ if applied in reverse, 

\[
\frac{|100\rangle_C+|010\rangle_C}{\sqrt2}|110\rangle_B
\xrightarrow{\tt UDD^{-1}}
|000\rangle_C|110\rangle_B,
\]
and similarly for the $|101\rangle_B$ and
$|011\rangle_B$ configurations. Thus, 
\begin{equation}
    |\phi_4\rangle=|000\rangle_C
    \frac{1}{\sqrt{3}}
    \left(
        |110\rangle_B|PPq\rangle_A+
        |101\rangle_B|PqP\rangle_A+
        |011\rangle_B|qPP\rangle_A
    \right).
\end{equation}
Finally, we can repeat the marking operation to return $B$ to its initial state 

\begin{equation}
    |\phi_5\rangle=|000\rangle_C|000\rangle_B
    \frac{1}{\sqrt{3}}
    \left(
        |PPq\rangle_A+
        |PqP\rangle_A+
        |qPP\rangle_A
    \right).
\end{equation}
Thus, the auxiliary registers are decoupled, and the particle registers contain the symmetric state of the three particles.

%%%%%%%%%%%%%%%%%%%%%%%%%%%%%%
\subsection{\tt{UDD}}
\label{subsubsec:UDD}
%%%%%%%%%%%%%%%%%%%%%%%%%%%%%%

Now, we illustrate how the \texttt{UDD} algorithm described in section~\ref{subsec:UDD} works.
We consider three qubits, with the $B$ register initialized in the state $\ket{110}_B$. Thus, the Hamming weight is $k=2$, and the occupied positions are $l_1=2$ and $l_2=3$.

To implement the algorithm, we introduce two auxiliary registers, $D$ and $E$, with $\lceil\log_2 n\rceil$ and
$n-1$ qubits, respectively. In our case, $n=3$, so both have 2 qubits.

As explained in section ~\ref{subsec:UDD}, in the second step, register $E$ is used to store the Hamming weight of $B$, which is the number of qubits in the $\ket{1}$ state. Since  $B=\ket{110}$,  $k$ equals 2 and therefore, $E$  rotates $\ket{00}$ to $\ket{10}$,

\begin{equation}
    \ket{\psi_2}=\ket{10}_E\ket{00}_D\ket{000}_C\ket{110}_B.
\end{equation}

In the third step, we prepare a superposition of the $k=2$ possible values of the counter $D$ (for this we need $E$ as an input). In this case, it is sufficient to apply a Hadamard gate to the first qubit of $D$, obtaining

\begin{equation}
    \ket{\psi_3}=\ket{10}_E\frac{\ket{00}_D+\ket{01}_D}{\sqrt{2}}\ket{000}_C\ket{110}_B.
\end{equation}

In the fourth step, we loop over the qubits of $B$, from $i=1$ to $i=3$. At each position, if $B_i=1$
and $D=0$, we apply a $\mathtt{X}$ gate to
$C_i$. Next, if $B_i=1$, we subtract one from the counter $D$, with modulo 4. Omitting the registers $E$ and $B$, which remain invariant, we start from
\begin{equation}
    \frac{\ket{00}_D\ket{000}_C+\ket{01}_D\ket{000}_C}{\sqrt{2}}.
\end{equation}
For $i=1$, as $B_1=0$, no operation is needed. 
For $i=2$, we have $B_2=1$. In the first term, $D=0$, so  $C_2$ is activated. In the second term, $D=1$, so $C$ remains unchanged. After subtracting one from $D$, we obtain

\begin{equation}
    \frac{\ket{11}_D\ket{010}_C+\ket{00}_D\ket{000}_C}{\sqrt{2}}.
\end{equation}
For $i=3$, we also have $B_3=1$. Now, the counter is zero in the second term, so $C_3$ is activated, while $C$ remains unchanged in the first term. After subtracting one from $D$, we have
\begin{equation}
    \frac{\ket{10}_D\ket{010}_C+\ket{11}_D\ket{100}_C}{\sqrt{2}}.
\end{equation}
Finally, we add $k=2$ to $D$, recovering its initial value,

\begin{equation}
    \ket{\psi_4}=\ket{10}_E\frac{\ket{00}_D\ket{010}_C+\ket{01}_D\ket{100}_C}{\sqrt{2}}\ket{110}_B.
\end{equation}
In the fifth step, we undo the operations used in the second step to calculate the Hamming weight in $B$, returning $E$ to its initial state. However, $D$ is entangled with $C$, so we have to apply the sixth step to decouple it.

In step 6.1 we use the marked position in $C$ to activate the corresponding qubits of $E$. To do this we use a cascade algorithm, we first activate $E_2$ if $C_3=1$ and then activate $E_1$ if $E_2=1$ or $C_2=1$. In the first term, $C=\ket{010}$, so we have $C_2=1$. Therefore $E_1$ is activated, giving $E=\ket{01}$. In the second term, $C_3=1$ so $E_2$ is activated first and followed by $E_1$, giving $E=\ket{11}$. Therefore, we obtain

\begin{equation}
    \ket{\psi_{6,1}}=\frac{1}{\sqrt{2}}\Big(\ket{01}_E\ket{00}_D\ket{010}_C+\ket{11}_E\ket{01}_D\ket{100}_C\Big)\ket{110}_B.
\end{equation}
In step 6.2, we subtract one from $D$ for each position where $B_i=E_i=1$. The first term remains invariant, while the second term has one subtracted, since $B_2=E_2=1$. Thus, both terms have $D=0$,
\begin{equation}
    \ket{\psi_{6,2}}=\ket{00}_D\frac{1}{\sqrt{2}}\Big(\ket{01}_E\ket{010}_C+\ket{11}_E\ket{100}_C\Big)\ket{110}_B.
\end{equation}
Finally, in step 6.3, we reverse the operations of step 6.1 to return $E$ to its initial state, 
\begin{equation}
    \ket{\psi_{6,3}}=\ket{00}_E\ket{00}_D\frac{\ket{010}_C+\ket{100}_C}{\sqrt{2}}\ket{110}_B.
\end{equation}
Thus, the auxiliary registers $D$ and $E$ are finally decoupled and ready for reuse, while $C$ contains a uniform superposition of the digits in $B$ , as desired. 

%%%%%%%%%%%%%%%%%%%%%%%%%%%%%%%
\section{Problems amenable to deployment of the algorithm}
\label{sec:problems}
%%%%%%%%%%%%%%%%%%%%%%%%%%%%%%%

The Complutense algorithm presented in this work is needed wherever the particle-register encoding \cite{Galvez-Viruet:2024hry} is used, because (anti)symmetrization enters the definition of creation and annihilation operators, and hence of every Hamiltonian written in second quantization. In this section we first recall how these operators are built and why this makes the algorithm (or a substitute thereof)  necessary for any problem formulated in second quantization. We then discuss two applications that we have already studied in this encoding: the fragmentation of jets into hadrons~\cite{Galvez-Viruet:2025rmy} and the equation of state of neutron stars~\cite{Castano-Garcia:2026oxg}.
\subsection{Second quantization in the particle-register encoding}
\label{Particle-register encoding}
Any discretized problem formulated in second quantization can be written in the particle-register encoding. To see why all of them require a symmetrization algorithm, recall how the encoding works.

The quantum memory is divided into $n$ registers of $n_{qb}+1$ qubits, each able to store a particle via a presence qubit, in $\ket{0}$ if the register is empty or $\ket{1}$ if it is occupied, and $n_{qb}$ qubits encoding the mode of the particle. Therefore, if we want to create a particle in a quantum memory with a single register, we must activate the presence qubit and write the corresponding mode in the remaining qubits,
\begin{equation}
    a_{\beta}^\dagger \equiv s^\dagger_{\beta}  \otimes C_{10} \ ,
\label{eq:adagger_individual_register}
\end{equation}
where $C_{ij}=\ket{i}\bra{j}$ are control operators acting on the presence qubit,\footnote{For $i\neq j$, $C_{ij}$ changes the presence qubit from $\ket{j}$ to $\ket{i}$, while $C_{00}$ and $C_{11}$ project onto empty and occupied registers, respectively.} $s^\dagger_\beta = \ket{\beta}\bra{0\ldots0}$, and $s_\beta = \ket{0\ldots0}\bra{\beta}$ are the set and scrap operators, which write the quantum numbers of mode $\beta$ into the selected register. 

The $a^\dagger_\beta$ operator acts on the empty register $\ket{\Omega}=\ket{0\ldots 0}\ket{0}$ as
\begin{equation}
    a_{\beta}^\dagger \ket{\Omega} = s^\dagger_{\beta} \ket{\underbrace{0\ldots0}_{n_{qb}}}\otimes C_{10}\ket{0} = \ket{\beta}\ket{1} \ ,
    \label{eq:adagger_individual_register_acting}
\end{equation}
whereas on an occupied register it returns zero, since $C_{10}\ket{1}=0$.

Multiple registers will be occupied in order, starting from the rightmost one. Thus, the field-theoretical creation (annihilation) operator is represented by the sum of such auxiliary operators acting on each register; any term that attempts to write a particle in an already occupied register (annihilate in an empty one) or that breaks the filling pattern, vanishes through the use of projectors. Thus, only the contribution of the operator acting on the appropriate register survives.

Finally, getting to the core discussion of this article, the correct statistics must be taken into account. If we create (annihilate) bosons, the corresponding symmetrizations must be effected, whereas for fermions we must antisymmetrize. 
Here our method shines because, if the memory on which the creation operator acts  already has the proper symmetrization, only the target register where a new particle is written has to be (anti)symmetrized with the already occupied ones.

Therefore, for bosons, the creation operator is given by
\begin{equation}
    a_\beta^{(n)\dagger} = \sum_{i=1}^{n}
    \underbrace{\rule[-2.7ex]{0pt}{0pt}
      \overbrace{\rule{0pt}{3.6ex}\Big[I^{(n-i)}\otimes\mathtt{S}^{\dagger}_{i}(\beta)\Big]}^{\substack{\text{Symmetrization from}\\ \text{$i-1$ to $i$ registers}}}
      \cdot
      \Big[\overbrace{\rule{0pt}{3.6ex}\sum^{i}_{k=1} \sqrt{k}\,P^{(n)}_i(k=n_\beta)}^{\substack{\text{Boson}\\ \text{enhancement}}}\cdot \overbrace{\rule{0pt}{3.6ex}P_0^{(n-i)}\otimes\big(s_\beta^\dagger\otimes C_{10}\big)_i\otimes P_{i-1}^{(i-1)}\Big]}^{\substack{\text{Creator of $\beta$}\\ \text{in the $i$-th register}}}
    }_{a_{\beta,i}^{(n)\dagger}} ,
    \label{eq:full_creation_operator}
\end{equation}
where the superscript $(n)$ denotes a memory of $n$ registers, $I^{(n)}$ is an identity on $n$ of them, and $P^{(n)}_i$ is the projector onto states of an $n$-register memory with registers $1$ to $i$ occupied and the rest empty. To account for boson enhancement, the operator $P^{(n)}_i(k=n_\beta)$ projects on memories in which exactly $k$ out of the $i$ occupied registers store mode $\beta$; and the corresponding state is therefore multiplied by $\sqrt{k}$. For fermions, the unitary step-symmetrizer $\mathtt{S}^{\dagger}_{i}(\beta)$ is replaced by the unitary step-antisymmetrizer $\mathtt{A}^{\dagger}_{i}(\beta)$, and the sum over $k$ collapses to the case $k=1$, which enforces the Pauli exclusion principle.

Note that Eq.~(\ref{eq:full_creation_operator}) is the sum of operators $a_{\beta,i}^{(n)\dagger}$ acting on different registers. In each of them, the filling pattern is enforced by $P_{i-1}^{(i-1)}$, which checks that there are no empty registers to the right of the $i$-th register, and $P_0^{(n-i)}$, which verifies that there are no occupied ones to the left. Thus, if these requirements are met, what remains is the creation of the particle in the $i$-th register via the single-register operator of Eq.~\eqref{eq:adagger_individual_register}, followed by its symmetrization with the $i-1$ particles already stored through $\mathtt{S}^{\dagger}_{i}(\beta)$. 

Notice that because of these projectors and $C_{10}$, the step-symmetrizer $\mathtt{S}^{\dagger}_{i}(\beta)$ is only applied to configurations with $i$ active registers where in the $i$-th register
the mode $\beta$
has just been stored. In this way, any other configuration makes the  state be annihilated before the symmetrization.

As we have seen, symmetrization is a fundamental step in the creation (annihilation) of particles, but it also enters every operation performed on the memory. For example, state preparation and every step of time evolution contain (anti)symmetrizations.

%%%%%%%%%%%%%%%%%%%%%%%%%%%%%%%
\subsection{Practical applications}

%%%%%%%%%%%%%%%%%%%%%%%%%%%%%%%
We now present a brief reminder of current problems of interest in hadron physics where lattice gauge theory struggles to provide first-principle results and quantum computing would be of much use. Therein our (anti)symmetrization method will find ample use when coding in a particle-register basis. We shall keep this discussion short and refer the reader to a review on the topic~\cite{Galvez-Viruet:2025ket}.

%%%%%%%%%%%%%%%%%%%%%%%%%%%%%%%
\subsubsection{Jet fragmentation into hadrons}
\label{subsec:fragmentation}
%%%%%%%%%%%%%%%%%%%%%%%%%%%%%%%
\begin{figure}[h]
    \centering
    \includegraphics[width=0.8\linewidth]{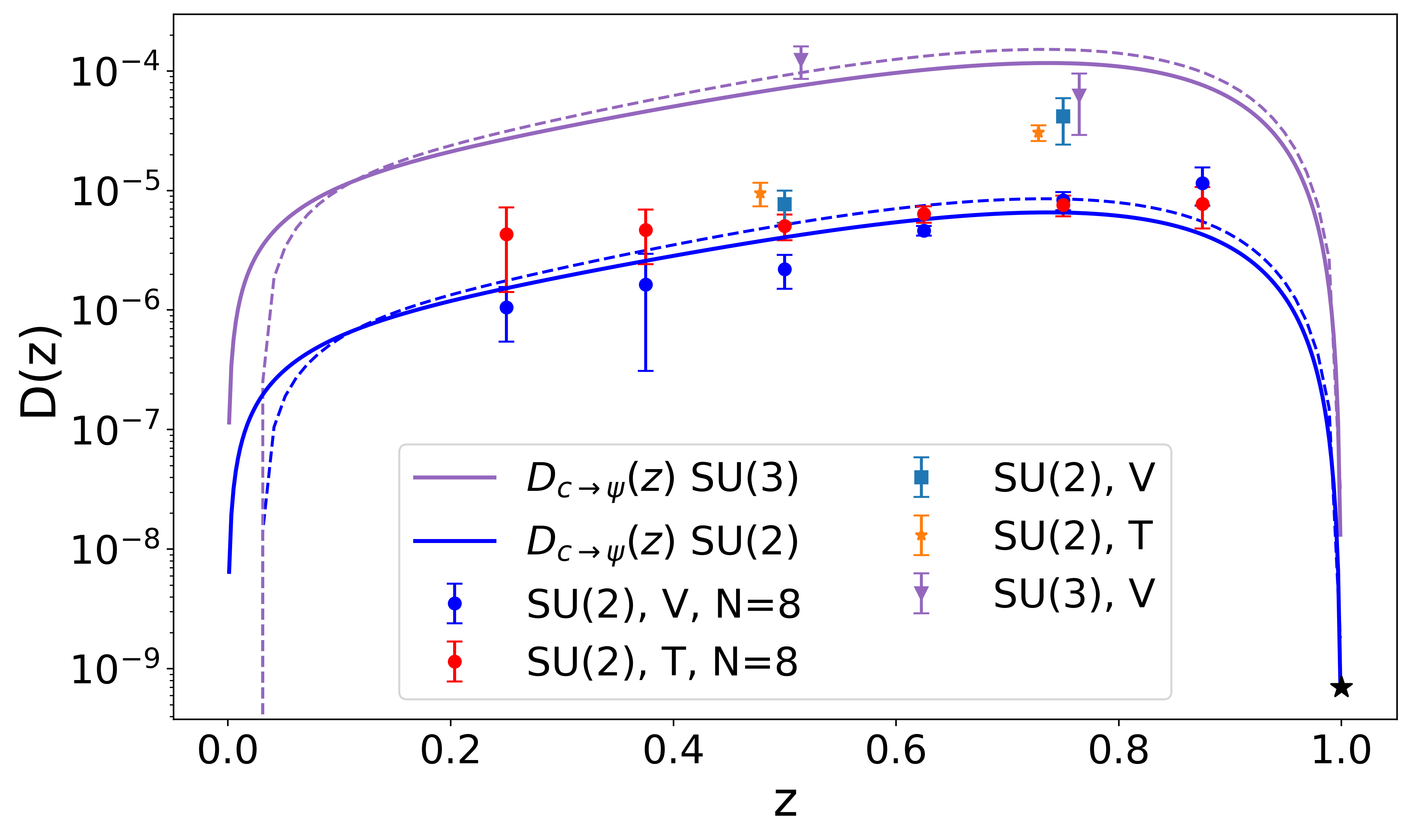}
    \caption{Evaluation of charm quark to $J/\psi$ fragmentation
function (data points) with an algorithm amenable for a quantum computer, compared to previous NRQCD computations (various lines). T: total QCD light-front Hamiltonian included. V: calculation where only the ``V'' potential terms are included~\cite{Galvez-Viruet:2026jgx}.}
    \label{fig:fragmentacion}
\end{figure}

Fragmentation functions require evaluating quark/gluon matrix elements taken on two different points in light-front time, so the states have to be evolved, with all semiinclusive identified hadron $h+X$ states contributing. Such computation escapes the possibilities of lattice gauge theory, but it would not be particularly difficult on a quantum computer. We in light-front QCD~\cite{Galvez-Viruet:2025rmy} and others~\cite{Li:2024nod} in the Nambu-Jona-Lasinio model have already explored the possibility of predicting fragmentation functions from first principles, see Fig.~\ref{fig:fragmentacion}. (Quantum computers can also be useful in the extraction of the fragmentation functions from data~\cite{Renteria-Estrada:2026rvw}.) With few quanta one can escape (anti)symmetrization, but as more particles intervene, the necessity to properly (anti)symmetrize (quarks) gluons becomes apparent.

In practical terms: when computing structure or fragmentation functions, our encoding is more efficient than Jordan-Wigner at high and intermediate $x$ or $z$, the longitudinal momentum fractions. At extremely low $x$, in the saturation regime where the number of particles approaches the number of quantum modes, the present particle-register encoding becomes subcompetitive respect to the Jordan-Wigner one (but can be used anyway if computational resources permit, there is no fundamental roadblock).

Many other applications in hadron structure, such as the computation of form factors and, more generically, the hadron tensor, are possible~\cite{Zou:2026cfk}, although for several of them lattice gauge theory is making strides~\cite{Lin:2026sxj}.

%%%%%%%%%%%%%%%%%%%%%%%%%%%%%%%
\subsubsection{Neutron star equation of state}
\label{subsec:nEoS}
%%%%%%%%%%%%%%%%%%%%%%%%%%%%%%%
The equation of state of neutron star cores is out of reach of lattice QCD because of the sign problem at finite chemical potential. In Ref.~\cite{Castano-Garcia:2026oxg} we showed that one first principles point $(\rho,P)$ at $\rho = 400 \text{MeV}/\text{fm}^3$, three times more precise than the present band, would reduce the uncertainty on the radius of a $1.4 M_{\odot}$ star from 36\% to 15\%. We then minimized the energy of Weyl-gauge QCD cells with a VQE with three quarks plus one gluon, or one quark plus two gluons. 

Under the $\sim 30$ qubit limit of the simulations, Ref.~\cite{Castano-Garcia:2026oxg} implemented calculation for two ansatzs:
\begin{enumerate}
    \item A color singlet superposition of three quarks with zero or one gluon ($3q+3q1g$).
    \item A color-triplet superposition of one quark with up to two gluons ($1q+1q1g+1q2g$).
\end{enumerate}
As explained in Sec.~\ref{Particle-register encoding}, particles are (anti)symmetrized in the moment they are created. The $3q+3q1g$ states should be antisymmetric under the interchange of quarks in the register, which is prepared creating particles one by one and iteratively antisymmetrizing them, whereas the $1q2g$ states should be symmetric under the exchange of the two gluons. 

The effect of these (anti)symmetrizers helps us to reduce the parameters of the VQE. If we want the variational method to give a reliable result, the trial states must be as general as possible while containing the minimum number of parameters for the program to converge. We must propose states that will not be annihilated by the particle register encoding due to incorrect symmetry in order not to waste parameters as well as introducing additional physical constraints. Moreover, keeping the registers properly ordered and symmetrized simplifies measuring the Hamiltonian and squared Gauss-law operator, $\mathcal{H}+\mathcal{G}^2$, cutting the number of Pauli terms of interaction terms of two gluons from $\mathcal{O}(10^6)$ to $\mathcal{O}(10^2)$. 

Overall, these three to four particle simulations already produce physical EoS curves and sound speeds, proving that the calculation works end to end. Scaling up to realistic cells (six or more quarks plus gluons, needing $\mathcal{O}(10^4)$ qubits) will require, apart of more powerful quantum computers, (anti)symmetrizing many more particles, making the improved algorithm directly relevant. 

%%%%%%%%%%%%%%%%%%%%%%%%%%%%%%%
\subsubsection{String-breaking and hadron production for Monte Carlo simulated showers}
\label{subsec:MonteCarlo}
%%%%%%%%%%%%%%%%%%%%%%%%%%%%%%%

A classical problem of high-energy physics is to compute the hadron showers produced from a collision among partons, and see which hadron species and with what momenta are produced, to calibrate experiment and eventually to test underlying theory~\cite{Sjostrand:2014zea}. 
However, the Monte Carlo programs which simulate these showers are based on classical simulations. Thus, when quantum effects become important, such as in baryon correlations where antisymmetry blocks part of the phase space, disagreement with experiment can become apparent~\cite{Graczykowski:2021vki}. While patches exist~\cite{Demazure:2022gfl}, a full quantum treatment of particle production from the QCD Hamiltonian will probably have to be addressed one day. We see no way of realistically attempting this than simulating the string-breaking process with a quantum computer~\cite{Bauer:2026rvd}.

%%%%%%%%%%%%%%%%%%%%%%%%%%%%%%%
\section{Outlook}
\label{sec:conclusions}
%%%%%%%%%%%%%%%%%%%%%%%%%%%%%%%

The resource analysis in Section \ref{apsec:SDDandS} distinguishes bosonic symmetrization from fermionic antisymmetrization. Sequential preparation scales quadratically in particle number up to logarithmic factors, with a contribution linear in the register width. The bosonic implementation additionally requires \texttt{UDD} to disentangle auxiliary records associated with repeated occupations, whereas the fermionic specialization avoids this step. This full-preparation scaling should be distinguished from the cost of inserting one mode into an already symmetrized state, which is smaller.

Compared with \cite{Abrams:1997gk}, our bounds replace the quadratic register-width dependence by a linear contribution, together with particle-number-dependent overheads. The sorting-based construction in \cite{Berry:2018ggo} achieves more favorable particle-number scaling for global preparation of distinct-mode configurations. Our distinguishing features are incremental insertion and explicit treatment of repeated bosonic occupations. Comparisons with \cite{Ward:2009woa} must additionally separate orbital loading and identification from exchange-symmetrization costs.

The appropriate comparison with \cite{Rule:2025ldr} is our fermionic specialization, which does not require \texttt{UDD}. The $T$-count reported and our elementary-gate count use different resource metrics. The nontrivial-state estimates include preparation costs absent from our basis-mode insertion problem. For repeated bosonic occupations, \cite{Liu:2024taj} instead prioritizes low-depth global preparation, including coherent superpositions of input lists. Our method therefore provides an efficient incremental primitive without implying a general resource advantage over global preparation algorithms. A more detailed gate-level decompositions of the sorting algorithms would be necessary to establish a meaningful comparison between the two approaches.

These computational considerations also motivate the potential applications to hadronic physics discussed below.

Given that this communication has been invited at a conference on the Physics of Excited Nucleons, we should quickly recapitulate how the computational methods here discussed might be of use in this subfield. In essence, we would deploy this work, given a powerful enough quantum computer, to predict observables there where Lattice methods may fall short. 
\begin{enumerate}
\item \emph{Very excited spectrum:} since exponentially suppressed signals have to be reconstructed from the raw numerical data, a variational approach based on a quantum computer may be an interesting complementary approach, perhaps along the lines of our earlier work~\cite{deArenaza:2024dhe} but with extensive variational bases and with the QCD Hamiltonian instead of a Cornell model. We do have an encoding of $H$ in the Weyl gauge at hand, as already discussed.
\item In high-energy production amplitudes and fragmentation, the Minkowski-based light-front approach on a quantum computer would become the go-to method if it could be deployed on a full quantum computer. This is probably the most direct and impactful application.
\item For hadron structure functions, Lattice methods have been devised~\cite{NieMiera:2026igj} that give access to certain aspects of parton distribution functions, but quantum computing could also make a contribution here and offer checks with different systematics.
\item The difficulties of working at finite density suggest the usage of quantum computers for nuclear matter computations and in neutron stars.
\end{enumerate}

%%%%%%%%%%%%%%%%%%%%%%%%%%%%%%%%%%%%%%%%%%%%%%%%%%%%%%%%
\section*{Acknowledgments}
%%%%%%%%%%%%%%%%%%%%%%%%%%%%%%%%%%%%%%%%%%%%%%%%%%%%%%%%

We acknowledge partial support from grants PID2025-170350NB-I00 and PID2025-169724NB-C21, 
 FPU21/04180, %<-- JJ scholarship
FPU24/00948 %<-- Nico scholarship
and FPU25/00872  %<-- Nahia scholarship
of the spanish government's program MICIU/AEI/10.13039/501100011033 and of ERDF/EU; and from the Horizon Europe programme (Grant Agreement No. 101292522, 
HADRON-2030).
%%%%%%%%%%%%%%%%%%%%%%%%%%%%%%%%%%%%%%%%%%%%%%%%%%%%%%%%
\appendix
%%%%%%%%%%%%%%%%%%%%%%%%%%%%%%%%%%%%%%%%%%%%%%%%%%%%%%%%
\newpage 

\section{Unit Adders and Unit Subtractors: implementation}
\label{sec:unit_adders_subtractors}
%%%%%%%%%%%%%%%%%%%%%%%%%%%%%%%%%%%%%%%%%%%%%%%%%%%%%%%%

\begin{figure}[h]
    \centering
    \begin{subfigure}[b]{0.5\linewidth}
        \centering
        \includegraphics[width=\linewidth]{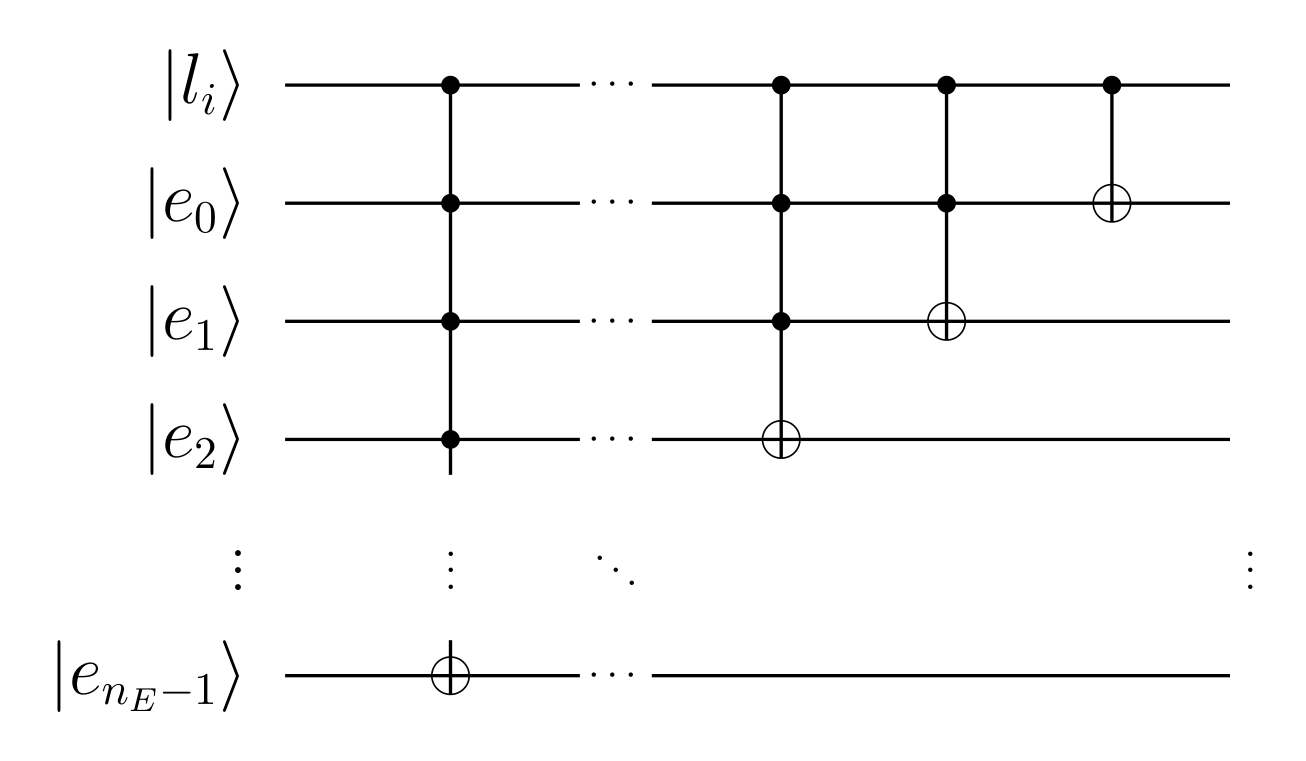}
        \caption{Unit adder.}
        \label{fig:unit_adder}
    \end{subfigure}\hfill
    \begin{subfigure}[b]{0.5\linewidth}
        \centering
        \includegraphics[width=\linewidth]{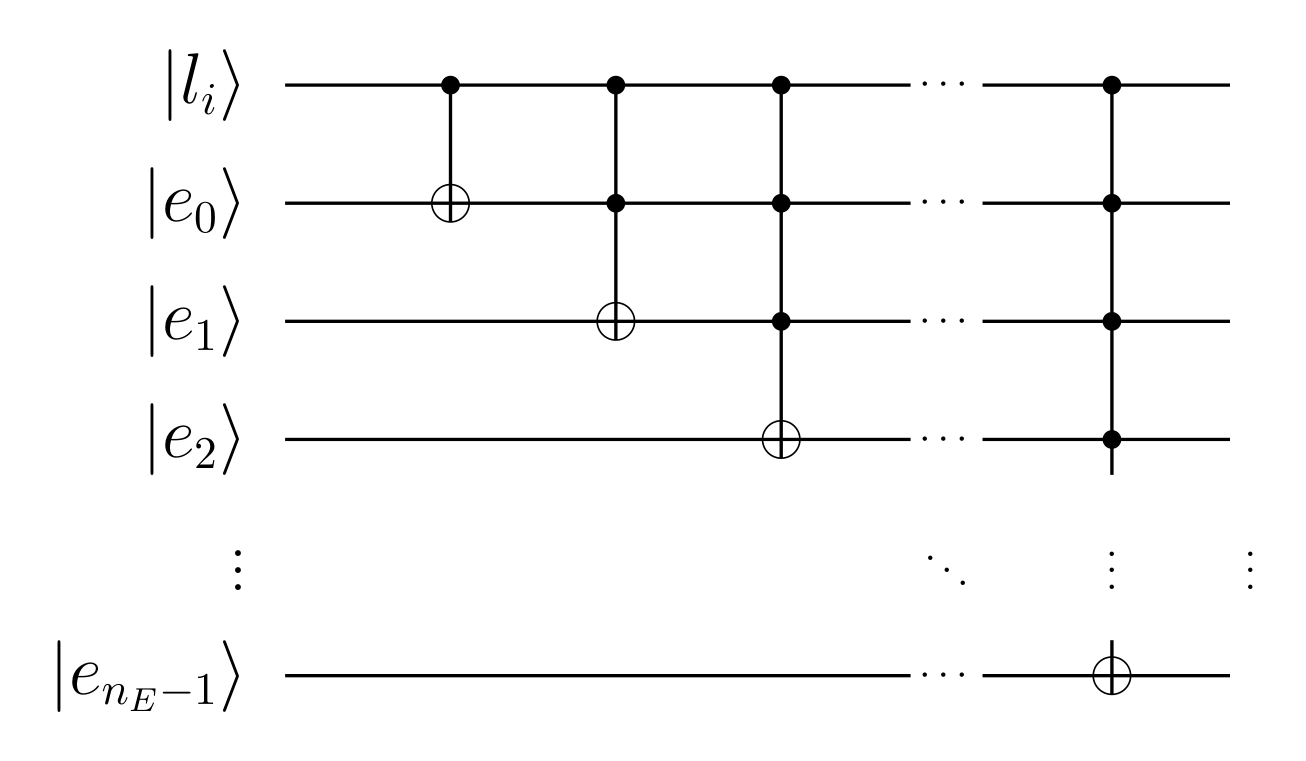}
        \caption{Unit subtractor.}
        \label{fig:unit_subtractor}
    \end{subfigure}
    \caption{Quantum circuits for the controlled unit arithmetic operations: (a) unit adder, and (b) unit subtractor.}
    \label{fig:unit_circuits}
\end{figure}

The goal of the second step of the UDD Algorithm (Section \ref{subsec:UDD}) is to evaluate the total number of 1s in a computational basis state $\ket{l}$ stored in a quantum register $B$ of $j$ qubits. To store this count, the algorithm requires an auxiliary register $E$ of $n_E$ qubits initialized to $\ket{0}^{\otimes n_E}$. This is achieved via Algorithm \ref{alg:UDD_Step_2}.

\begin{algorithm}[htbp]
\caption{Step 2 of the UDD Algorithm (Section \ref{subsec:UDD})}\label{alg:UDD_Step_2}
\begin{algorithmic}[1]
\Require Register $B$ in basis state $\ket{l}$, auxiliary register $E$ in state $\ket{0}^{\otimes n_E}$
\Ensure Total number of 1s in $l$ stored in register $E$

\For{$i\gets0$ \textbf{to} $n_{\mathrm{qb}}-1$}
    \State Apply unit adder to $E$ if $l_i=1$
\EndFor
\State \Return Register $E$
\end{algorithmic}
\end{algorithm}

Expressed in terms of its constituent bits, $l=l_{n_{\mathrm{qb}}-1} \dots l_1 l_0$ with $l_i\in \{0,1\}$, the required operation of the unit adder is the following:
\begin{itemize}
    \item If $l_i=1\implies \vert{}k\rangle_E \to \vert{}k + 1\pmod{2^{n_E}}\rangle_E$ (adds it to register $E$)
    \item If $l_i=0\implies \vert{}k\rangle_E \to \vert{}k\rangle_E$ (it does nothing)
\end{itemize}
That is, if $l_i=1$, the integer stored in $E$ is incremented by 1 modulo $2^{n_E}$; otherwise, the identity operation is performed. This means that the unit adder needs to control that $\ket{l_i}$ is in state $\ket{1}$ for it to act over register $E$. The only thing left to explain is how the addition itself is implemented in $E$. 

Decomposing the state of register $E$ into its individual bits $e_i\in \{0,1\}$, when adding 1 to a binary number, a bit only changes if all of the bits to its right are 1: 
\begin{itemize}
    \item Bit $e_0$ always changes.
    \item Bit $e_1$ changes only if $e_0=1$.
    \item Bit $e_2$ changes only if $e_0=1$ and $e_1=1$.\\
    \vdots
    \item Bit $e_i$ changes only if $e_0=1$, $e_1=1$, \ldots, $e_{i-1}=1$.
\end{itemize}
To do this on our quantum circuit, we note the following: if we modified $e_0$ first, we would destroy the carry information needed to determine whether $e_1$ and $e_2$ should change. Therefore, the circuit is applied from highest to lowest weight (and, as we said before, conditioning each gate on the input bit $l_i$). The resulting implementation is shown in Figure \ref{fig:unit_adder}.

The unit subtractor performs the inverse operation of the unit adder:
\begin{equation}
\ket{l_i}_B \ket{k}_E \mapsto \ket{l_i}_B \ket{k-l_i \pmod{2^{n_E}}}_E.
\end{equation}
Because this operation is the exact inverse, it is implemented by executing the sequence of gates of the unit adder in reverse order (starting from the least significant bit $e_0$ up to the most significant bit $e_{n_E - 1}$), as depicted in Figure \ref{fig:unit_subtractor}. This implementation is simple but costly in terms of gate executions, requiring $\cO(n_E^2)$ single-qubit gates and $\tt CNOT$s. More efficient algorithms achieve linear scaling, see \cite{Gidney:2018haa}.

%%%%%%%%%%%%%%%%%%%%%%%%%%%%%%%

\end{document}